\documentclass[graybox]{svmult}

\usepackage{type1cm}        
\usepackage{makeidx}         
\usepackage{graphicx}        
\usepackage{multicol}        
\usepackage[bottom]{footmisc}

\usepackage{newtxtext}       %
\usepackage{newtxmath}       

\usepackage{todonotes}
\usepackage{subfigure}
\usepackage{lipsum}  
\usepackage[hidelinks]{hyperref}

\makeindex             

\begin{document}

\title*{Experiences in Using the V-Model as a Framework for Applied Doctoral Research}
\author{Rodrigo Falc\~{a}o, Andreas Jedlitschka, Frank Elberzhager, and Dieter Rombach}
\institute{Rodrigo Falc\~{a}o,  Andreas Jedlitschka, and Frank Elberzhager  \at Fraunhofer IESE, Fraunhofer-Platz 1, 67663 Kaiserslautern (Germany) \\ \email{{rodrigo.falcao, andreas.jedlitschka, frank.elberzhager}@iese.fraunhofer.de}
\and Dieter Rombach \at Professor Emeritus at RPTU Kaiserslautern, Postfach 3049, 67663 Kaiserslautern (Germany) \\ \email{dieter.rombach@siak-kl.com}}
%
%
\maketitle

\abstract*{The pervasive role played by software in virtually all industries has fostered ever-increasing development of applied research in software engineering. In this chapter, we contribute our experience in using the V-Model as a framework for teaching how to conduct applied research in empirical software engineering. The foundational idea of using the V-Model is presented, and guidance for using it to frame the research is provided. Furthermore, we show how the framework has been instantiated throughout nearly two decades of PhD theses done at the University of Kaiserslautern (RPTU Kaiserslautern) in partnership with Fraunhofer IESE, including the most frequent usage patterns, how the different empirical methods fit into the framework, and the lessons we have learned from this experience.}

\abstract{The pervasive role played by software in virtually all industries has fostered ever-increasing development of applied research in software engineering. In this chapter, we contribute our experience in using the V-Model as a framework for teaching how to conduct applied research in empirical software engineering. The foundational idea of using the V-Model is presented, and guidance for using it to frame the research is provided. Furthermore, we show how the framework has been instantiated throughout nearly two decades of PhD theses done at the University of Kaiserslautern (RPTU Kaiserslautern) in partnership with Fraunhofer IESE, including the most frequent usage patterns, how the different empirical methods fit into the framework, and the lessons we have learned from this experience.}

\section{Introduction}
\label{emse24:sec:intro}

In recent decades, we have witnessed the rise and consolidation of empiricism as a key aspect of software engineering research. With its roots in early (and rare) initiatives between the mid-1960s and mid-1970s, empirical software engineering (ESE) has become an established and relevant field for researchers and practitioners. Over time, the teaching of ESE has found its way into universities, especially in graduate programs.

Education in empirical software engineering in graduate courses often focuses on the careful teaching of empirical strategies. For example, when the focus of a given course is on controlled experiments, its syllabus will comprise the introduction of basic concepts such as the meaning and purpose of experimental studies, variables, hypothesis testing, types of experimental design, the experimental procedure, and the application of statistical methods.

However, conducting graduate-level research requires students to have the ability to not only master empirical strategies. Empiricism is part of a broader endeavor, namely research itself, which demands identification of an adequate problem, declaration of a research question, definition of general and specific goals, and so forth. Structuring software engineering research that uses the empirical approach and navigating through its phases is not trivial, especially in PhD-level research, where more rigor is required in all phases of research and a multi-method strategy is often needed. Furthermore, conducting and supporting the development of software engineering PhD theses in an applied research setting contributes another variable to the structuring of the research, for the research must address not only a scientifically important problem but also one with practical relevance.

In this chapter, we report on how the V-Model can guide teaching and performing applied research using the empirical approach. The V-Model is a conceptual model that provides a framework for a certain set of elements in a given context. These elements are organized in the shape of a ``V'', where the first element takes place at the top-left point of the shape. It is followed by other elements along the left side until the bottom point and then by the last elements along the right side of the shape. The raison d'être for the shape of the V-Model is that the elements featured on the right side map back to elements on the left side. In RPTU Kaiserslautern, the V-Model has been used as \textit{a virtual model for products and for empiricism}. The virtual product model abstracts from what concrete processes have in common to fit any process; likewise, the virtual empirical model creates a frame that abstracts from concrete empirical strategies.

This chapter is the result of nearly 20 years of using the approach in the context of PhD theses done at RPTU Kaiserslautern in cooperation with Fraunhofer IESE\footnote{The list of books in the series “PhD Theses in Experimental Software Engineering”, whose review is the main source for this chapter, can be found at \url{https://s.fhg.de/phd-theses-in-ESE}.}. Approximately half of the theses published between 2005 and 2023 used the V-Model explicitly (and an even higher number used it as a reference model, although it may not have been made explicit in some theses). To prepare this chapter, we reviewed these theses (23 in total), extracting usage patterns of the V-Model. Furthermore, we invited the authors of these theses to share the lessons they learned using the V-Model in their research. The feedback we received (from twelve authors) has been integrated into this text.

The remainder of this chapter is organized as follows: Section~\ref{emse24:sec:background} describes the origins of the V-Model and where the ideas to adapt it to the classroom came from; Section~\ref{emse24:sec:vmodel-approach} introduces the framework; Section~\ref{emse24:sec:usage-patterns} discusses the main variations in which the framework has been used; Section~\ref{emse24:sec:lessons-learned} reflects on the lessons we have learned; and Section~\ref{emse24:sec:further-reading} provides further reading recommendations.

\section{Background}
\label{emse24:sec:background}

The V-Model has its origins in the software engineering field, but was neither initially intended to support applied research nor related to ESE. We have used the V-Model for this purpose as we found it an adequate vessel to bring our experience in applied research to the classroom.

\subsection{The origins of the V-Model}\label{emse24:sec:origins}

\begin{figure}[h!]
\centering
\includegraphics[scale=.65]{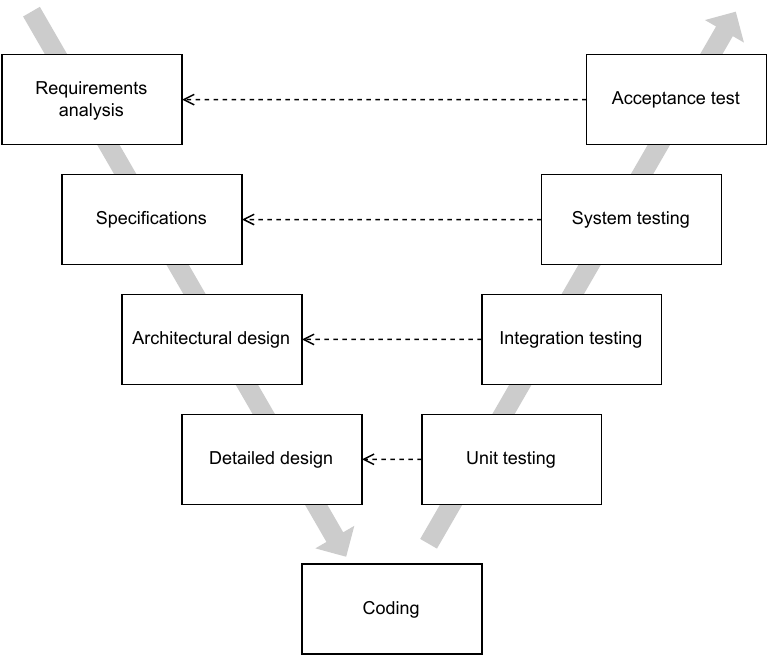}
%
%
\caption{A variant of the original V-Model being used to frame a traditional software development lifecycle (adapted from \cite{mathur2010advancements}).}
\label{fig:original-v-model}       
\end{figure}

The introduction of the V-Model dates back to 1986 and was first featured in the Software Engineering Journal \cite{rook1986controlling} in the context of the software testing process. It extends the waterfall process, relating requirements and design activities on the left side to validation activities on the right side. Figure~\ref{fig:original-v-model} illustrates one of many variations of the original V-Model.

\subsection{Bringing the applied research approach to the classroom}\label{emse24:sec:qip}

While the V-Model was originally introduced as a \textit{process model}, this is not the only way it can be used. In the early 90's the forth author introduced the V-Model as a \textit{product model} and, shortly later, as an \textit{empirical model} while designing a software engineering program focused on processes and empiricism at RPTU Kaiserslautern. When used as a product model, the V-Model is explicitly non-sequential. The idea was that any product being built through any process (be it sequential or iterative) should be consistent with the model. Therefore, the V-Model as a product model is a process-independent tool useful for building products.

When used as an empirical model (which has been the case in the PhD theses and, therefore, is on focus in this chapter), the V-Model is useful for process engineering, i.e., engineers benefit from its results. The principles behind the empirical V-Model have been used previously in applied research projects and are still being used now. At Fraunhofer IESE, the empirical research process is described as a cycle \cite{jedlitschka2013empirical} with the following steps: \textit{Characterize problem}, \textit{Set research goals}, \textit{Choose process for implementing the strategy}, \textit{Execute the research as planned}, \textit{Analyze the results}, and \textit{Package the results}. The principles underlying the process have been brought to the classroom with two adaptations. The first was that while it may be sufficient in certain industry projects to characterize the practical problem, in the academic environment the scientific problem should always be evident. The second was that the process was designed as a cycle, implying a continuous improvement process, whereas in the educational context, a well-defined scoping strategy that explicitly points to an end yields greater benefits. For this purpose, the V-Model has been introduced as an instrument to bring the principles of the empirical approach to applied research to the lectures and dissertations in the graduate courses of RPTU Kaiserslautern. 

\subparagraph{Tip: Focusing on principles}
    When using the V-Model as a product model, the focus is on providing students with stable principles. These principles should be valid independent of the process to be used. Processes can change, and the way you build software varies, but in the end, there will be a clear problem, a set of requirements, a design, a solution, etc., and it does not matter whether these were generated sequentially or iteratively. This stability of principles ensures the reliability of the V-Model as a framework for teaching software engineering processes. The same is valid for using the V-Model as a framework for using the empirical approach in applied research in software engineering, independent of the empirical methods used.

\section{The V-Model framing approach}
\label{emse24:sec:vmodel-approach}

\begin{figure}[h!]
\sidecaption
\includegraphics[scale=.65]{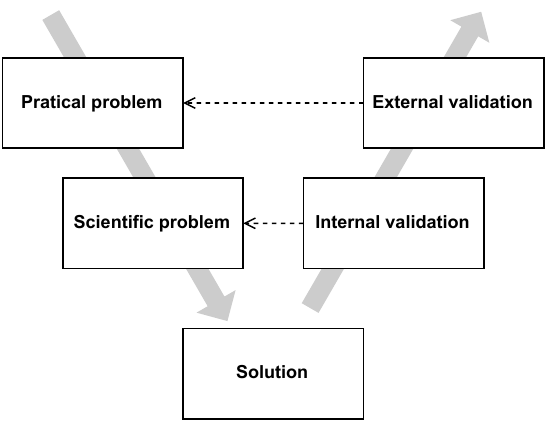}
%
%
\caption{The meta-model of the empirical V-Model.}
\label{fig:v-meta-model}       
\end{figure}

To frame applied software engineering research with the support of empirical methods, we have, in most cases, used a 5-stage V-Model. The starting point is the \textit{practical problem}, followed by the \textit{scientific problem}, the \textit{solution}, the \textit{internal validation}, and the \textit{external validation}. The meta-model of the empirical V-Model is illustrated in Figure~\ref{fig:v-meta-model}.

%

\subparagraph{Tip: Pre-requisites}
    While, in principle, there is no particular pre-requisite to understanding the V-Model as a framework for ESE, it might make sense to introduce it in the context of an ESE course, where students should become familiar with the empirical approach in software engineering, because putting the V-Model framing approach in practice will require previous knowledge on, at least, the major empirical methods (case studies, surveys, and controlled experiments).

\subsection{The practical problem}\label{emse24:sec:practical-problem}

In an applied research setting, the research shall address real/practical problems, i.e., a problem that is known and experienced by practitioners. The practical problem -- also referred to in the V-Model as ``industry problem'' or ``state of the practice'' -- is usually illustrated through an example. In fact, from our experience, this is the typical starting point: describing one or more real scenarios where the problem is contextualized and explained. While we may start from a scenario description, at the end we need one or more clear and concise \textit{problem statements}. Therefore, the general question at this stage in the V-Model is:

\paragraph{\textbf{What is the practical problem faced by practitioners?}}

At times, the practical problem may include one or more hypotheses about a potential improvement. For instance, in a given context, the problem may be the high costs of keeping software architecture documentation up-to-date. One may ask what could be done to decrease the maintenance effort, for example, by 20\%\footnote{Quantifiable improvement goals are usually preferable when it comes to the hypotheses related to the practical problem, because companies want to know whether it is worthwhile to invest in a certain change.}. At this point, an implicit hypothesis about a potential solution (which is still unknown and, therefore, to be devised) is that such a solution would achieve the improvement goal. At other times, there might be implicit hypotheses about the problem statement. Is the problem indeed a problem, or would it be someone's biased opinion? These are things to be investigated empirically.

\begin{svgraybox}

\textit{Example: Elicitation of context-aware functionalities\footnotemark}

Context-aware functionalities are functionalities that consider context to produce a certain system behavior, typically an adaptation or a recommendation. They are a desired feature in many software-based systems. However, they are hard to discover because they deal with implicit input. This is even more evident when trying to discover \textit{unexpected} context-aware functionalities that aim to delight users by making the application exhibit what can be called ``smart behaviors''. However, the very fact that these functionalities are unexpected implies that they are not trivial to identify. In a survey with practitioners, the researchers found that although some context modeling techniques exist, practitioners do not use them to support the elicitation of context-aware functionalities. Practitioners perceive context modeling as a complex activity and have therefore been overlooking it. It is particularly hard to figure out which contextual factors -- individually or in combination with each other -- are relevant for a given user task of interest. As a result, opportunities for describing and implementing context-aware functionalities are missed.

\end{svgraybox}

\footnotetext{The running example used throughout this chapter refers to \cite{falcao2023data}.}

\paragraph{\textbf{Empiricism for investigating practical problems}}

The investigation of the practical problem may serve different purposes:

\begin{itemize}
    \item \textbf{Providing evidence for the practical problem:} While the practical problem may sound reasonable when described in a certain scenario with a proper example, an empirical investigation can provide evidence about the existence and relevance of the problem. As elaborated by Jedlitschka and Pfahl \cite{jedlitschka2005reporting} in their guidelines for reporting controlled experiments, it is important to declare what the problem is, where it occurs, who has observed it, and why it is important to be solved.
    \item \textbf{Exploring details of the problem:} Empirical studies can be used to test hypotheses about the problem and help pinpoint the root causes of the problem from a practical point of view. Furthermore, empirical investigation can help understand the size of the problem and measure its impact.
    \item \textbf{Explaining the causes of the problem:} When the researchers have hypotheses about the causes of the problem, these can be examined using empirical methods.
    
\end{itemize}

In our experience, one of the most frequent empirical methods employed to investigate the practical problem is \textbf{surveys with practitioners} (examples of such usage can be found in \cite{abukwaik2018proactive, falcao2023data, rost2019task, trendowicz2008software}). Surveys can be implemented in different ways. For instance, when the aim is to investigate problems from a general population, surveys are frequently implemented using online questionnaires as the data collection strategy, as this can help to reach more potential participants. On the other hand, when the research is done in the context of a certain company, expert interviews are a common option (e.g., \cite{menzel2012optimizing}). For a comprehensive introduction to surveys in software engineering, we refer to a series of articles by Kitchenham and Pfleeger \cite{kitchenham2002principles}.

Besides surveys, \textbf{case studies} are another typical empirical method used to investigate the practical problem (an example can be found in \cite{knodel2011sustainable}). Depending on the research context, one or more case studies may be conducted to draw conclusions about the practical problem. Guidelines on case studies in software engineering are provided by Runeson et al. \cite{runeson2012case}.

Even if the empirical investigation of the practical problem is done in the context of one single company, the risk of the investigated problem being only relevant in the context of the given company should not jeopardize the research, for the next step in the V-Model is to identify the scientific problem that underlies the practical problem.

\subsection{The scientific problem}\label{emse24:sec:scientific-problem}

While the practical problem refers to the state of the practice, the scientific problem is concerned with the state of the art. The goal at this stage in the V-Model is to identify the scientific problem (also referred to as ``research problem'' or ``research challenge'') that underlies the practical problem. The general  question between the identification of the practical problem and the identification of the scientific problem is:


\paragraph{\textbf{How does the state of the art address the problem?}}

If an adequate solution already exists in the state of the art, the immediate follow-up question to be answered is \textit{Why has the existing state-of-the-art solution not been used?}. There may be a transfer issue. It may be the case, for example, that practitioners lack proper guidance, methods, or tools, to benefit from the state-of-the-art solutions. Conversely, if an adequate solution is not yet available, the gap in the state of the art becomes self-evident.

Based on the problem, one option is to explore the solution space, looking for existing alternatives with no assumptions about potential solution ideas. Existing solutions, when found, are accompanied by their limitations, which may provide hints for uncovering the scientific problem. Alternatively, it is possible to explore the solution space not only based on the problem but also with one implicit hypothesis (or maybe more) about a candidate solution: Why has \textit{a certain approach} not been used to address the problem? Or: \textit{Has a certain approach been used?} The investigation may lead to finding a complete absence of evidence about the application of the approach or to identifying the limitations found by other researchers when trying to use the intended approach.

\begin{svgraybox}

\textit{Example: Discovering state-of-the-art context modeling approaches}

The researchers involved in the previous example had the idea of automating context modeling using a \textit{data-driven approach}. So they reviewed the literature to search for existing solutions that followed the same idea. Although some approaches exist, they do not provide a comprehensive end-to-end solution to automate context modeling for eliciting context-aware functionalities. Among other things, the researchers found that, in some cases, no process or conceptual model is provided. In other cases, no context model is explicitly generated; the solution is coupled to a particular data analysis algorithm or context model; or the ultimate goal of the approach is not the same. In other words, there is a lack of systematic end-to-end approaches for data-driven context modeling for the elicitation of context-aware functionalities.
\end{svgraybox}

\paragraph{\textbf{Empiricism for investigating scientific problems}}

The investigation of the scientific problem aims at:

\begin{itemize}
    \item \textbf{Declaring the scientific problem:} The goal is to pinpoint the concrete, well-defined scientific problem that underlies the practical problem of interest.
    \item \textbf{Providing evidence for the relevance of the scientific problem:} The scientific problem must be worth being investigated and must therefore be an open problem (i.e., not yet solved).
    \item\textbf{Explaining the root causes of the problem:} When the researchers have hypotheses about the root causes of the scientific problem, they can investigate them empirically.
    \item\textbf{Exploring potential solution strategies:} Previous work can provide directions about both promising and discouraging research directions in terms of solutions.
 
\end{itemize}

\textbf{Systematic literature reviews} (SLRs) -- as proposed by Kitchenham and Charters \cite{kitchenham2007guidelines} -- have been used frequently as a way to gather evidence about scientific problems in software engineering. Being not a primary study type, systematic literature reviews can accumulate both positive and negative evidence (including, but not limited to, empirical evidence) about existing scientific problems, where positive evidence refers to the previous acknowledgment and documentation of these problems and negative evidence means the absence of reports in the literature about a particular or any general solution for the problems. Besides SLRs, mapping studies are also useful to review the state of the art. Guidelines for this strategy have been proposed by Petersen et al. \cite{petersen2008systematic}.

Sometimes, the starting point may be a certain fact about a phenomenon observed in practice. For example, a certain software development process \textit{X} may be found to be more efficient than an alternative software development process \textit{Y} in Company A; however, in Company B, process \textit{Y} is found to be more efficient than \textit{X}. If the fact can be observed but the reason is unclear, explanatory research can be carried out to search for cause-effect relationships that might explain the root cause of the problem. In such cases, one or more \textbf{controlled experiments} can be used to support the investigation.

Once the scientific problem has been identified, the next step is to devise a solution for the problem.

\subsection{The solution}\label{emse24:sec:solution}

At the bottom of the V-Model, the design and implementation of a solution for the problem is a creative activity undertaken by the researchers. The guiding questions between the scientific problem and the solution are:

\paragraph{\textbf{What is the research goal and what solution strategy can be used to achieve it?}}

The characteristics of the solution will be specific to the problem under investigation. The solution may contribute to the foundations of the software engineering field to which it belongs (by providing and/or organizing concepts, for example), to the methods used in the field (by describing processes or providing guidance), and/or to the technology used in the field (through the implementation of tools/automation), for example. In any case, at this point, a solution is planned and implemented, be it a model, a method, a tool, etc., or a combination of them. 


\begin{svgraybox}
    \textit{Example: Developing a data-driven context modeling framework}
    
    In the absense of a comprehensive approach, the data-driven context modeling idea was refined and, raising the level of abstraction, the researchers decided to describe and implement a \textit{framework}. The framework comprises a conceptual model explaining entities and relationships in the domain of data-driven context modeling, and a process providing step-by-step guidance for what must be done. The framework itself was designed to be decoupled from any particular implementation of the contextual model or the data analysis algorithm. In order to apply the data-driven approach, the framework was instantiated through the development of a concrete context meta-model and a concrete data analysis algorithm.
\end{svgraybox}

\paragraph{\textbf{Empiricism while creating the solution}}

In our experience, empirical methods are not typically \textit{applied} at the solution stage (although there have been exceptions). However, \textit{empirical work} does happen at the solution stage as it anticipates the planning of the next stages (internal and external validation). Empiricism at the solution stage means:

\begin{itemize}
    \item \textbf{Defining hypotheses about the solution:} The solution is expected to improve in some way the as-is situation described through the practical and the scientific problems.
    \item \textbf{Outlining the evaluation:} The hypotheses must be verified using one or more empirical methods, which should be chosen in advance.
    \item \textbf{Performing pre-evaluation of the solution:} Before bringing the solution to the formal validation stages, preliminary validation studies can be performed with the support of empirical methods to provide early feedback and help improve the shape of the final solution.
\end{itemize}

In the ideal case, hypotheses about the solution should be specified right after the definition of the solution idea, for the implementation of the solution idea should be driven (and eventually checked) through the lens of the improvement goals implied (or explicitly mentioned) in the hypotheses.

Having defined the hypotheses and the solution strategy, the evaluation plan starts to take shape. Empirical methods for evaluating different aspects of the solution concepts in different contexts from different perspectives should be considered and chosen according to the needs and constraints of the research. From our experience, in most cases, the detailed plan for the internal and external validations can wait until the next stages (internal validation and external validation) to be formulated. In some cases, however, an initial evaluation is performed during the solution stage. It can take a variety of forms, ranging from a series of less formal feedback-gathering exercises such as an \textbf{early application with industry practitioners} (see \cite{naab2012enhancing} for an example) to \textbf{pilot tests} (``try runs''). In the latter case, a complete experimental procedure may be prepared in advance: through pilot tests, the procedure can be evaluated and adapted for the experiment to be carried out in the next stages. Furthermore, pilots help understand how the solution is perceived by others, even if it is still a partial solution.

Regardless of whether a preliminary evaluation has been performed on the solution or not, the hypotheses about the solution should be evaluated in the next stages, starting with an internal validation.

\subsection{Internal validation}\label{emse24:sec:internal-validation}

The first step on the right side of the V-Model is internal validation, sometimes referred to as \textit{academic validation} or \textit{scientific benefit}. At this stage, the hypotheses about the solution concerning the scientific problem are tested. This means that internal validation is concerned with checking the hypotheses about the qualities of the solution to address the scientific problem. The guiding question for internal validation is:


\paragraph{\textbf{What benefits does the solution offer in terms of addressing the scientific problem?}}

Put differently, at this stage, the researcher wants to verify how (or if) the solution improves the state of the art. The benefits to be observed can be derived from the hypotheses defined right after the definition of the solution idea, as mentioned previously. Note, however, that the initial hypotheses may change after the solution is ready, for the very process of implementing the solution can shed new light on the benefits that were not completely understood. 

\begin{svgraybox}
    \textit{Example: Validating the automated context model through an experiment}

    The researchers collected contextual data from a real mobile app to apply the implementation of the framework. The tool analyzed the data and automatically generated a context model that revealed, based on the contextual data, how different contexts have influenced a certain user task of interest. Then a controlled experiment was designed to evaluate whether or to which extent the resulting context model (primary factor) can help individuals identify relevant contexts to describe context-aware functionalities. Furthermore, the subjects of the treatment group (i.e., participants who used the context model) also answered a debriefing survey to report on the usability of the context model to support their elicitation task.
\end{svgraybox}

\paragraph{\textbf{Empiricism during internal validation}}

Empiricism is at the core of internal validation. Internal validation is about applying one or more empirical methods to test the hypotheses about the benefits of the implemented solution to address the scientific problem. Empiricism at this stage includes:

\begin{itemize}
    \item \textbf{Defining/refining hypotheses:} After implementing the solution, the researcher is ready to define/review the initial hypotheses and check whether they can be refined or whether new hypotheses should actually be considered.
    \item \textbf{Designing empirical evaluation:} If the evaluation procedure was not designed in the late phase of the solution stage, it has to be defined in the early stages of the internal evaluation.
    \item \textbf{Performing empirical evaluation and reporting on it:} The chosen empirical strategy is carried out, data is collected and analyzed, and the results are documented.
\end{itemize}

According to our experience, \textbf{controlled experiments} have been the most popular empirical strategy used to perform internal validation (comprehensive information about how to conduct controlled experiments in software engineering is organized by Wohlin et al. in \cite{wohlin2012experimentation}; detailed information on how to report controlled experiments is provided by Jedlitschka et al. in \cite{jedlitschka2007reporting}). In some cases, internal validation was performed with graduate students of RPTU Kaiserslautern and, in other cases, with professionals at Fraunhofer IESE. In any case, even when professionals participate, the evaluation bears the character of an internal validation because it is not performed in a natural (i.e., industrial) setting where people are experiencing the problem under investigation. However, it is unquestionable that when internal validation is performed with the participation of professionals, they contribute much more practical expertise to the results than students.

Controlled experiments are frequently accompanied by a \textbf{survey} organized into two parts: the characterization of the participants (sometimes this is a separate survey sent in advance to help block participants and ensure a balanced distribution of the subjects in the control and treatment groups), and the collection of post-experiment impressions (debriefing survey). The survey data can shed light on the quantitative analysis that follows the execution of the controlled experiment by providing qualitative and (additional) quantitative data to the researchers. As regards qualitative analysis, a comprehensive introduction on using qualitative methods in ESE is provided by Seaman \cite{seaman1999qualitative}. \textbf{Simulation-based studies} (SBS) can also be used to validate solution proposals. An introduction to SBS can be found in \cite{defranca2020role}. Finally, in our experience, \textbf{case studies} have also been occasionally used to perform internal validation (examples can be found in \cite{jung2019context, rudolph2020generation}).

\subsection{External validation}\label{emse24:sec:external-validation}

Finally, at the final stage of the V-Model, the researchers perform external validation. This stage is also called \textit{industrial validation} or \textit{practical benefit}. The point now is to check whether practitioners can benefit from the solution in their natural settings to solve their real problems. The guiding question at this stage is:

\paragraph{\textbf{To what extent is the solution beneficial in practice?}}

While a controlled experiment in an internal validation may provide evidence that a certain factor (namely: the solution itself or something derived from the solution) can be identified as the cause of measurable improvement in the current situation, this does not necessarily imply that the solution is \textit{practical}, i.e., beneficial to practitioners. External validation aims to strengthen the validation of the solution in a practical setting, which can be supported by empiricism.

\begin{svgraybox}
    \textit{Example: Using the data-driven modeling approach in a company}
    
    A medium-sized car-sharing company joined the research and the solution was used in a real setting. They have a mobile app that their users use to manage the booking of cars. The agile team responsible for creating and maintaining the app was introduced to the framework and the implementation usage. The team followed the process and used the tool to create context models for different user tasks of interest, and then used the models to support creative elicitation sessions within the team in order to identify interesting context-aware functionalities to improve their system. At the end of a period of applying the framework, the researchers interviewed the team to get their impressions of the practical benefit (as well as difficulties) experienced by them.
\end{svgraybox}

\paragraph{\textbf{Empiricism during external validation}}

    This stage is similar to the previous one (internal validation), which is why the role of empiricism is similar: It is about \textbf{defining/refining hypotheses} and \textbf{designing}, \textbf{performing}, and \textbf{reporting on the empirical evaluation}. The fundamental differences are the audience (the evaluation occurs in an industrial environment), the sample size, and, typically, the instrument.

    Similar to the practical problem stage, \textbf{case studies} and \textbf{surveys with practitioners} are the usual empirical strategies for evaluating the solution in a practical setting.

\section{Execution flows and further usage patterns}
\label{emse24:sec:usage-patterns}

In the previous section, we presented the most common way in which we have experienced the usage of the V-Model to frame applied research in software engineering using empirical strategies. Although the stages' logical order -- practical problem, scientific problem, solution, internal validation, and external validation -- suggests a fairly sequential process, it is important to emphasize that the approach does not describe a process but a framework where varying execution flows can occur. Besides that, we have also experienced alternative usage patterns of the V-Model framing approach in several PhD theses done at Fraunhofer IESE and RPTU Kaiserslautern over the years. This section explores those usage patterns.

\subsection{Process execution flows}

While it is possible to follow the stages of the V-Model strictly in sequence, variations and/or iterations are often observed. The most frequent variations are:

\begin{figure}[h]
 \centering
 \subfigure[Iterations between problem and solution.]{\includegraphics[width=0.4\textwidth]{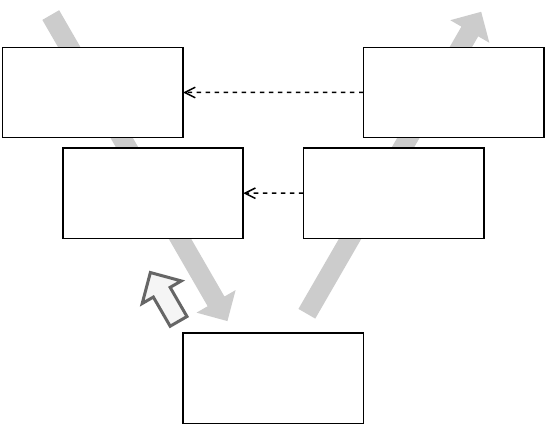}\label{fig:iteration-problem-solution}}
 \hspace{0.5cm}
 \subfigure[Iterations between problem and validation.]{\includegraphics[width=0.4\textwidth]{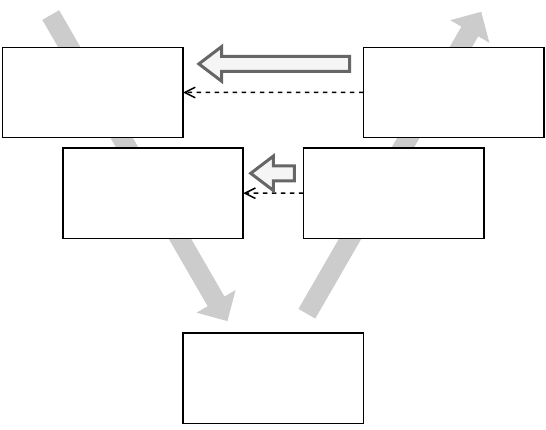}\label{fig:iteration-problem-validation}}
 \subfigure[Inverse validation.]{\includegraphics[width=0.4\textwidth]{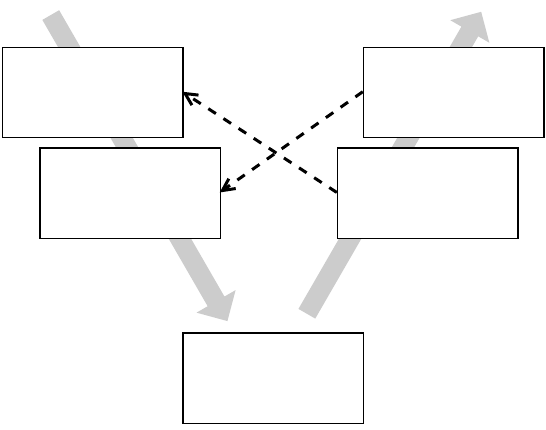}\label{fig:crossed-validation}}
 \hspace{0.5cm}
 \subfigure[Example of research without external validation.]{\includegraphics[width=0.4\textwidth]{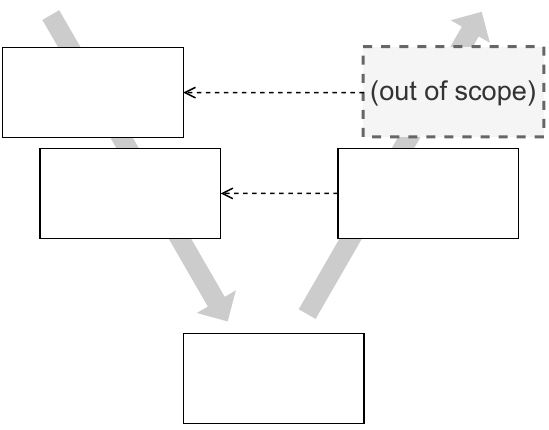}\label{fig:no-external-validation}}
 \caption{Typical variations in the execution flow of the V-Model.}
 \label{fig:multi-figure}
\end{figure}

\begin{itemize}
    \item \textbf{Multiple iterations between problem(s) and solution definition:} The definition of a solution idea does not begin after the scientific problem is identified. On the contrary, it is natural that potential research ideas, or at least research directions, emerge when a given problem is under investigation. Therefore, as the practical problem and the scientific problem are concretely defined, an early solution idea is refined until it is designed and implemented. See Figure~\ref{fig:iteration-problem-solution}.

    \item \textbf{Iterations between problems and validations:} During the definition of the practical problem, hypotheses about the future solution can be defined that are expected to be tested, in principle, during external validation. Likewise, when the scientific problem is known, hypotheses about the solution can be identified, and these are expected to be tested, in principle, in the internal validation stage. However, as mentioned in Sections~\ref{emse24:sec:solution} and~\ref{emse24:sec:internal-validation}, developing the solution may lead to refining the hypotheses and/or introducing new ones. These therefore need to be double-checked in light of the practical and scientific problems. See Figure~\ref{fig:iteration-problem-validation}.

    \item \textbf{Inverse validation:} While it has been common usage that the internal (academic) validation tests hypotheses related to the scientific problem and the external (industry) validation tests those related to the practical problem, sometimes the academic validation (implemented, for example, through a controlled experiment) tests hypotheses related to the practical problem, and the industry validation (implemented, for example, through a case study) tests hypotheses of the scientific problem. This can happen for different reasons, such as time and scope constraints of the project where the research has been developed and intrinsic characteristics of the problem under investigation. See Figure~\ref{fig:crossed-validation}.

    \item \textbf{Either internal \textit{or} external validation:} The reasons mentioned in the previous item also cause another effect in the process execution flow of the V-Model: Depending on the nature of the problem and/or the scope and time constraints of the graduate course, either internal or external validation will be performed (by implementing one or more empirical strategies). One implication is that the test of some declared hypotheses is left out of the scope of some theses. For example, in \cite{rost2019task, falcao2023data}, case studies for external validation were envisioned but not executed within the scope of the theses; in \cite{muller2007analyzing}, no internal validation was performed, but rather a series of case studies in an industrial setting. See Figure~\ref{fig:no-external-validation}.

\end{itemize}

\subsection{Further usage patterns}

In most cases, a 5-stage V-Model as illustrated in Section~\ref{emse24:sec:vmodel-approach} has been used to frame applied research in empirical software engineering. Nonetheless, alternative patterns have emerged from many theses over the years.

\begin{itemize}
    \item \textbf{Multiple instances of the V-Model:} When PhD research tackles more than one practical problem, each of them with one individual underlying scientific problem, it may make sense to have multiple V-Models to frame the work: one for each practical problem. One example can be found in the PhD thesis of Zhang \cite{zhang2015vital}.
    
    \item \textbf{Research approach and validation approach:} Nearly half of the PhD theses we reviewed used the V-Model to frame the overall research approach or method, usually in the early chapters of the thesis. In some other cases, however, the V-Model was placed in the late chapters related to validation. In yet other cases, the V-Model was used to structure the chapters of the thesis (e.g., \cite{menzel2012optimizing, ocampo2009remis})
    
    \item \textbf{Positioning the hypotheses:} Some theses position the hypotheses on the left side of the V-Model, in connection with the practical and scientific problems. This resembles the original application of the V-Model in the ESE lectures at RPTU Kaiserslautern in the early 2000s. Back then, the left side described the problems with associated hypotheses, which should guide the research; the right side indicated the empirical strategies (often controlled experiments) used to address the left-side hypotheses\footnote{A copy of the slide set used in the early 2000s at RPTU Kaiserslautern (formerly University of Kaiserslautern) can be found at \url{https://zenodo.org/doi/10.5281/zenodo.11544897}.}. Other theses leverage the structure of the V-Model to visually place their hypotheses on different parts of it (e.g., on the solution stage, on one of the validation stages, or even on arrows linking the validation stages to the problem stages -- see one example in \cite{john2010pattern}).
    
    \item \textbf{N-stage V-Model:} In some theses, the V-Model does not have five stages. Interesting examples include the usage of a 6-stage V-Model as in \cite{anastasopoulos2014evolution}, where the author features three problem stages on the left side and the corresponding three benefit stages on the right side (the bottom stage is not shown, but ``solution'' is implied there). Similarly, in another thesis \cite{armbrust2010scope}, the author used a 7-stage V-Model, featuring on the right side the stages ``validation: controlled experiment (university context)'', ``validation: industry case studies'', and ``roll-out: application in projects''.

\end{itemize}

Despite the variants we found in using the V-Model, the essence of the framing approach remains almost unchanged: In the end, it is all about finding a practical problem, identifying the underlying scientific problem, developing a solution, and validating it internally and externally.

\section{Lessons learned}
\label{emse24:sec:lessons-learned}

\begin{figure}
        \centering
        \includegraphics[scale=0.65]{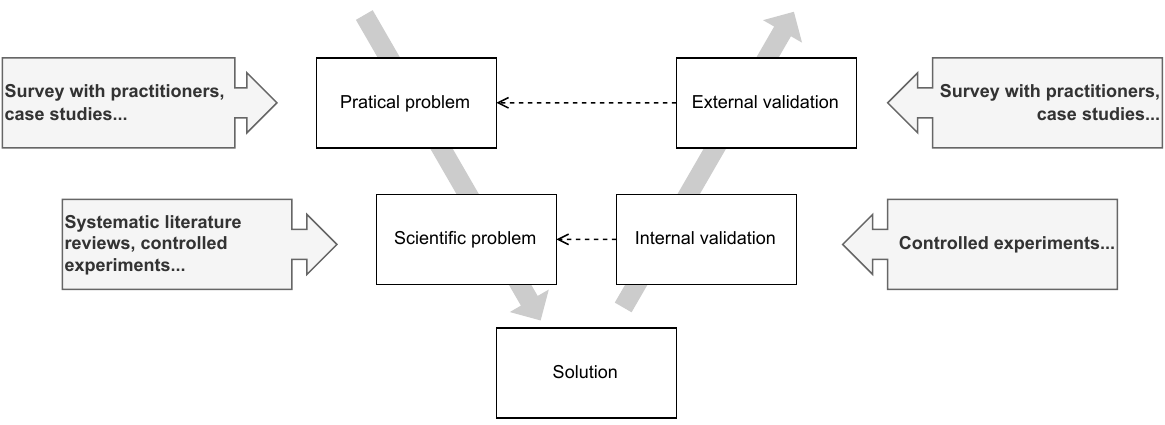}
        \caption{Summary of the typical empirical strategies used in the stages of the V-Model framework.}
        \label{fig:v-model-with-strategies}
    \end{figure}

According to our experience, using the V-Model as a framework for teaching ESE at RPTU Kaiserslautern, the origins of which date back to the early 2000s, has evolved in different ways. In the early days, controlled experiments were the empirical strategy in focus; later, the framework was used to support the structuring of multi-strategy research. Empiricism has been used not only to validate the solution but also to characterize the problem. Figure~\ref{fig:v-model-with-strategies} summarizes the typical empirical strategies applied to each stage of the V-Model.

The various usages that we have experienced over the years also reflect the intended, but perhaps not self-evident, flexibility of the framework. The V-Model helps guide the research process, which is especially beneficial for PhD-level students, where there is frequent uncertainty about what should be done to achieve the final goal. Yet the V-Model should not constrain the research but rather be adapted to the research constraints and goals.

Besides providing guidance along the research process, another feature of the V-Model that has proven beneficial is that the transitions between the stages function as checkpoints, acting as quality gates and monitoring the research progress. While PhD-level research is often solitary work, at these checkpoints, supervisors should contribute their experience to the work by reviewing the results.

\subparagraph{Feedback from a survey with former PhD students}

As we were interested in how former PhD students experienced the V-model during their PhD, we prepared four simple questions to obtain some lessons learned. As mentioned in the introduction, we asked 23 people whose dissertations we had identified as explicit uses of the V-model. We received responses from twelve authors. These have been aggregated into the summary presented in the following.

\paragraph{\textbf{``Why did you use the V-model?''}}

The majority answered that the model was presented and given by the leading PhD supervisor. The model was well-established at a certain point and therefore the standard procedure for making progress in one's PhD. As many other PhD students also followed the model and regular discussions and exchange took place among all these students, one learned how others applied the model, saw the advantages, and could reflect on and adapt this for one's own PhD. The model was often considered as a systematic and structured way, and it was considered logical to use the model.

\paragraph{\textbf{``How did you use the V-Model? Was it clear for you how the V-Model should or could be used to frame your PhD research?''}}

The model was sometimes considered as some kind of project plan, i.e., it helped structure the concrete steps of how to proceed and provided guidance on what to do in which order. This means that the students understood how the different steps in the V-model are aligned and what to do in which order. Therefore, it was also considered as some kind of template, which sometimes was even explicitly covered in the outline of the thesis and the concrete headings. For most, the model became clear (at least after some time) and was then also used implicitly to some extent.

\paragraph{\textbf{``What went well in following the V-Model approach?''}}

 One common answer was that the process felt natural and intuitive, was easy to follow and logical. It helped in organizing the work and the written thesis. The good guidance of the model itself together with the discussions helped to structure the thesis. Feedback could be given easily this way, too. This also meant that the right questions were asked at the right time, implying that a logical order of the steps in the PhD was enforced by following the V-model. Certain phases, such as problem identification or evaluation, were sometimes emphasized and it was mentioned that the model provides support to address these phases adequately. Finally, the overall alignment of the five steps was rated positively.

\paragraph{\textbf{``Were there drawbacks or shortcomings in using the V-Model? What could be improved?''}}

 The most general comment was that some details in the model remain unclear, as there is only little detail available about the individual phases. This was mitigated to a certain extent during the discussions with other PhD students and the leading supervisor. Consequently, the idea was sometimes mentioned that the process should be describe in more detail. Some comments address specific steps. For example, it was mentioned that better support in defining the problem could be helpful. Another example was the validation phase where some hints, especially regarding validation with external partners, could improve the evaluation. The evaluation as such might be too small to be generalizable and the model does not prevent this risk. Also, the relation between the practical problem and the scientific problem was sometimes difficult to draw by some PhD students. Another statement was that the V-model seems to be more accurate for practical PhD theses than for more theoretical theses.

\begin{svgraybox}
    
\textit{Summarizing the feedback}
\begin{itemize}
    
    \item The V-Model has been perceived as an intuitive and useful tool to guide PhD candidates in structuring their research.
    \item The guidance provided by the approach has helped both the planning of concrete steps and the writing of the theses.
    \item Students have been able to adapt the V-Model to their own needs.
    \item The V-Model facilitates feedback giving.
    \item The approach fits better with applied doctoral research.
    \item Defining the problem, particularly the scientific problem, is sometimes perceived as a challenging step.

\end{itemize}
\end{svgraybox}

\subparagraph{Tip: The identification of the problem}
In our experience, defining the practical problem and especially identifying the underlying scientific problem is challenging for students. While we acknowledge this difficulty, we see it as a natural part of the applied doctoral process. The identification of a relevant scientific problem, taking a practical problem as a starting point, requires creativity and effort from the candidate to explore the problem space.

\section{Further reading}
\label{emse24:sec:further-reading}

\begin{figure}
        \centering
        \includegraphics[scale=0.65]{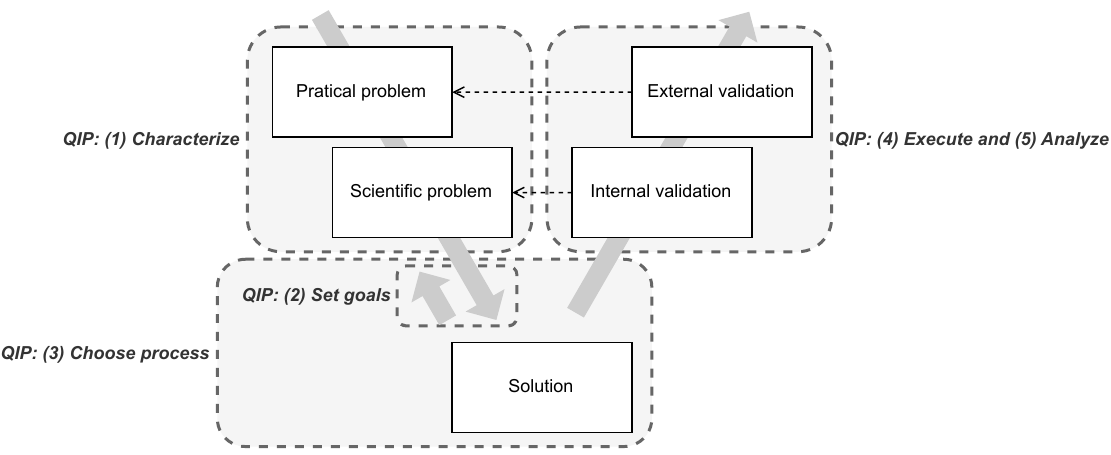}
        \caption{Mapping between the steps of the QIP and the V-Model research meta-model.}
        \label{fig:v-model-qip}
    \end{figure}

The principles behind using the V-Model to teach ESE come from the empirical research approach applied at Fraunhofer IESE, which is based on the Quality Improvement Paradigm (QIP). As the QIP has been the reference model adopted by ISERN\footnote{ISERN (International Software Engineering Research Network) is the international network of ESE research, aimed at supporting international collaboration on empirical software engineering. \url{https://isern.iese.de/}}, its principles are well known by the ESE community. Therefore, for a process framework for planning, executing, and analyzing individual empirical studies of any type, we refer to the QIP \cite{basili1994experience}, as described in \cite{jedlitschka2013empirical}. Figure~\ref{fig:v-model-qip} illustrates how the QIP steps map to the V-Model\footnote{The step \textit{Package (6)} refers to the documentation, reporting, and dissemination of the research results. Since the V-Model is used to frame multi-method research, this is a cross-cutting step across the stages. For this reason, it has been omitted in the mapping.}.

While we contribute a framework for the usage of multiple empirical strategies in applied research in this chapter, the empirical strategies themselves have only been mentioned briefly. For a comprehensive introduction to the major empirical strategies in software engineering, we refer to Wohlin et al. \cite{wohlin2012experimentation}. Advanced aspects of empirical strategies are provided by Shull et al. \cite{shull2007guide}. For further information about contemporary empirical methods, including case surveys and simulations in empirical software engineering, we refer to Felderer and Travassos \cite{felderer2020contemporary}.

For the operationalization of goals in all types of studies, we refer to the Goal, Question, Metric (GQM) approach \cite{basili1994gqm} and in particular the GQM goal template for formulating goals in a formal way.

The V-Model framework is useful not only in the educational context but also in any applied research setting. In that sense, it is not the only existing framework. For example, an alternative for a multi-method strategy in applied research using ESE is presented by Runeson et al. \cite{runeson2020design}. In their work, the authors use the Design Science paradigm for framing empirical research, focusing on ``articulating and communicating prescriptive software engineering research contributions''. Another alternative has been proposed by Gorschek et al. \cite{gorschek2006model}. The authors present a ``research approach and technology transfer model'' described in a series of steps that sometimes occur in industry (e.g., ``problem/issue'', ``dynamic validation'') and other times in academia (e.g., ``study state of the art'', ``validation in academia'').

\begin{acknowledgement}
We acknowledge the direct and indirect contributions of many colleagues and former colleagues at Fraunhofer IESE, the majority of whom were former PhD candidates themselves. They have significantly contributed to building this body of knowledge by applying the V-Model framing approach in various ways over the years.
\end{acknowledgement}
\section*{Appendix}
\addcontentsline{toc}{section}{Appendix}
A slide set to support teaching the V-Model framing approach for applied research in ESE that is based on the content of this chapter is available at \url{https://zenodo.org/doi/10.5281/zenodo.11544897} and can be reused by those who may be interested in it.


%
%
\bibliographystyle{abbrv}
\bibliography{refs}

\end{document}